# Deriving Two Sets of Bounds of Moran's Index by Conditional Extremum Method

Yanguang Chen

(Department of Geography, College of Urban and Environmental Sciences, Peking University, 100871, Beijing, China. Email: chenyg@pku.edu.cn)

**Abstract:** Moran's index is a basic measure of spatial autocorrelation, which has been applied to varied fields of both natural and social sciences. A good measure should have clear boundary values or critical value. However, for Moran's index, both boundary values and critical value are controversial. In this paper, a novel method is proposed to derive the boundary values of Moran's index. The key lies in finding conditional extremum based on quadratic form of defining Moran's index. As a result, two sets of boundary values are derived naturally for Moran's index. One is determined by the eigenvalues of spatial weight matrix ($n\lambda_{\min} \leq$ Moran's $I \leq n\lambda_{\max}$), and the other is determined by the quadratic form of spatial autocorrelation coefficient (-1< Moran's $I$ <1). The intersection of these two sets of boundary values gives four possible numerical ranges of Moran's index. A conclusion can be reached that the bounds of Moran's index is determined by size vector and spatial weight matrix, and the basic boundary values are -1 and 1. The eigenvalues of spatial weight matrix represent the maximum extension length of the eigenvector axes of *n* geographical elements at different directions. This work solves one of the fundamental problems of spatial autocorrelation analysis.

# 1 Introduction

Moran's index is one of basic and important statistical measure of spatial autocorrelation. One of main functions of Moran's index is to judge whether or not there is redundant information in a spatial sample. Moran's index was created by analogy with the autocorrelation coefficient of time series analysis (Moran, 1950; Haggett *et al*, 1977; Odland, 1988). The temporal autocorrelation coefficient actually proceeded from Pearson correlation coefficient. For Pearson correlation coefficient and temporal autocorrelation coefficient, the boundary values are -1 and 1. For Pearson correlation, if two variables are equal to or perfectly collinear with each other, the correlation coefficient reaches its extreme values, 1 or -1. For temporal autocorrelation coefficient, if time lag is zero, the autocorrelation coefficient reaches its upper limit 1. In short, whether it is a simple correlation coefficient or an autocorrelation coefficient, the direct correlation between a variable and itself constitutes the ceiling and floor of the correlation coefficient value. The upper limit value (ceiling) is 1 and the lower limit value (floor) is -1. Moran's index is actually proposed to measure the indirectly correlation between a geographical element and itself, and this correlation is caused by the interaction based on spatial network. In this case, the absolute value of Moran's index cannot reaches the value of 1. Therefore, the boundary values of Moran's index are regarded as -1 and 1.

However, there are different views on the boundary value of the Moran's index in the field of geography. It has been found that, in some cases, the upper limit and lower limit of Moran's index are determined by the maximum and minimum eigenvalue of spatial weight matrix (de Jong *et al*, 1984; Tiefelsdorf and Boots, 1995; Xu, 2021). In theory, for *n* spatial elements, Moran's index values comes between *n* times of minimum eigenvalue and *n* times of maximum eigenvalue of the globally normalized spatial weight matrix. This result can be derived by the principle of Rayleigh quotient (Xu, 2021). The question now is whether the original boundary values, namely -1 and 1, should be denied? Things are not so simple. There is no reason to suggest that these two sets of boundary values are incompatible with each other. In fact, many parameters or measures have different boundary values under different conditions. For example, fractal dimension bears at least three sets of boundary values (Addison, 1997; Chen, 2011; Feder, 1988; Mandelbrot, 1983).

Unfortunately, for a long time, there has been controversy about the boundary values of Moran's index, and there is no consensus. Due to misunderstanding of Moran's boundary values, some absurd calculations of Moran's index (that is, the absolute values of Moran's *I* are greater than 1) have also been accepted by geographers. The purpose of this paper is to theoretically solve the boundary value problem of the Moran's index using a novel method. Two sets of boundary values of Moran's index will be derived, and they are both proved to be valid under different conditions. The relationships between two sets of boundary values can be clarified by mathematical analysis.

The rest parts are organized as follows. In Section 2, two sets of boundary values are derived. Based on quadratic form of Moran's index, the special boundary values are derived. Based on the conditional extreme values of the quadratic form, the general boundary values and special boundary values are simultaneously derived. In Section 3, an empirical analysis are made to verify the mathematical derivation results. In Section 4, several related questions are discussed, and finally, in Section 5, the discussion will be concluded by summarizing the main points of this work.

## 2 Theoretical results

### 2.1 Boundary values based on quadratic form

Spatial autocorrelation measure is based on distance matrix or generalized distance matrix. Distance matrix is always symmetric, and asymmetry is a special case. A real symmetric matrix is one type of Hermitian matric in linear algebra. For a real symmetric matrix **W** and a real vector **z**, we can construct a real quadratic form based on $n$ elements such as

$$I = \mathbf{z}^T \mathbf{W} \mathbf{z}, \tag{1}$$

where $I$ denotes Moran's index, **z** is a standardized vector, **W** is a globally normalized weight matrix, and T refers to matrix transpose. We can find an orthogonal matrix **C**, which satisfies a linear transformation as follows

$$\mathbf{z} = \mathbf{C}\mathbf{y}, \tag{2}$$

in which **y** denotes a new vector. Thus, Moran's index can be expressed as

$$I = \mathbf{z}^T \mathbf{W} \mathbf{z} = (\mathbf{C}\mathbf{y})^T \mathbf{W} \mathbf{C}\mathbf{y} = \mathbf{y}^T (\mathbf{C}^T \mathbf{W} \mathbf{C})\mathbf{y} = \mathbf{y}^T \Lambda \mathbf{y} = \sum_{i=1}^{n} \lambda_i y_i^2, \tag{3}$$

where $\lambda_i$ refers to the $i$th eigenvalue of **W**, $\Lambda = \mathbf{C}^T \mathbf{W} \mathbf{C}$ is a diagonal matrix composed of these eigenvalues. According to the nature of orthogonal matrix, $\mathbf{C}^{-1} = \mathbf{C}^T$, that is, the inverse matrix of **C** is equal to the transpose matrix of **C**. So we have

$$\mathbf{C}^T \mathbf{C} = \mathbf{E}, \tag{4}$$

where **E** is an identity matrix. So, from equation (2) it follows

$$\mathbf{z}^T \mathbf{z} = (\mathbf{C}\mathbf{y})^T \mathbf{C}\mathbf{y} = \mathbf{y}^T (\mathbf{C}^T \mathbf{C})\mathbf{y} = \mathbf{y}^T \mathbf{y} = n. \tag{5}$$

In equation (3), if $\lambda_i = \lambda_{\max}$, we have the maximum Moran's index, $I_{\max} = n\lambda_{\max} \geq 0$; In contrast, if $\lambda_i = \lambda_{\min}$, we have the minimum Moran's index, $I_{\min} = n\lambda_{\min} \leq 0$. Therefore, we obtain the following relations

$$n\lambda_{\min} = \lambda_{\min}(\mathbf{y}^T\mathbf{y}) \leq \mathbf{z}^T\mathbf{W}\mathbf{z} = \sum_{i=1}^{n}\lambda_i y_i^2 \leq \lambda_{\max}(\mathbf{y}^T\mathbf{y}) = n\lambda_{\max}. \tag{6}$$

From equations (5) and (6), we can derive

$$\lambda_{\min} \leq \frac{\mathbf{z}^T\mathbf{W}\mathbf{z}}{\mathbf{z}^T\mathbf{z}} \leq \lambda_{\max}, \tag{7}$$

which can be derived from the theorem of Rayleigh quotient (Xu, 2021). This suggests one set of boundary values $I \geq n\lambda_{\min}$, and $I \leq n\lambda_{\max}$. That is, the range of values for the Moran's index is

$$n\lambda_{\min} \leq I \leq n\lambda_{\max}, \tag{8}$$

where, in theory, $\lambda_{\min} \leq 0$ and $\lambda_{\max} \geq 0$, but in reality, $\lambda_{\min} < 0$ and $\lambda_{\max} > 0$.

## 2.2 Boundary values based on conditional extremum

As indicated above, Moran's index can be expressed as a real quadratic form. Based on quadratic form, the boundary values of Moran's index can be derived in a novel way. According to the principle of linear algebra, for a real quadratic form, the rule of taking differentiation of equation (1) are as follows

$$\frac{\partial I}{\partial \mathbf{z}} = 2\mathbf{z}^T\mathbf{W}, \tag{9}$$

$$\frac{\partial y}{\partial \mathbf{z}^T} = 2\mathbf{W}\mathbf{z}. \tag{10}$$

Suppose that Moran's index bear extreme values, including the maximum value and minimum value. Assuming that Moran's index tends to the maximum value being subject to $\mathbf{z}^T\mathbf{z}=n$, we have a simple programming model as below

$$\begin{cases} \max & I = \mathbf{z}^T\mathbf{W}\mathbf{z} \\ \text{s.t.} & \mathbf{z}^T\mathbf{z} = n \end{cases}, \tag{11}$$

where *max* refers to "maximizing", and *s.t.* denotes "subject to". Apparently, the constraint condition is a very simple quadratic form, which is based on an identity matrix, that is $n = \mathbf{z}^T\mathbf{E}\mathbf{z} = \mathbf{z}^T\mathbf{z}$. To find the extremes of Moran's index, a Lagrange function can be constructed as follows

$$L(\mathbf{z}) = \mathbf{z}^T\mathbf{W}\mathbf{z} + \lambda(n - \mathbf{z}^T\mathbf{E}\mathbf{z}), \tag{12}$$

where $\lambda$ is Lagrange multiplier. Taking derivative of equation (12) with respect to $\mathbf{z}$ and $\mathbf{z}^T$, respectively, yields

$$\frac{\partial L(\mathbf{z})}{\partial \mathbf{z}} = 2\mathbf{z}^T\mathbf{W} - 2\lambda\mathbf{z}^T\mathbf{E} = 2\mathbf{z}^T(\mathbf{W} - \lambda\mathbf{E}) = \mathbf{O}^T, \tag{13}$$

$$\frac{\partial L(\mathbf{z}^T)}{\partial \mathbf{z}^T} = 2\mathbf{W}\mathbf{z} - 2\lambda\mathbf{E}\mathbf{z} = 2(\mathbf{W} - \lambda\mathbf{E})\mathbf{z} = \mathbf{O}, \tag{14}$$

where $\mathbf{O}$ denotes zero vector. Equations (9) and (10) were made use in the above process of taking derivative.

It must be clear that $\mathbf{z}$ is a non-zero real vector. In this case, we have two sets of solutions to equation (13) or equation (14). The two sets of solutions results in two sets of boundary values for Moran's index. Case 1, for any standardized nonzero real vector $\mathbf{z}$, finding a spatial weight matrix to meet equations (13) and (14). Case 2: for any globally normalized spatial weight matrix, finding a group of eigenvectors to satisfy equations (13) and (14). Based on case 1, general boundary values of Moran's index can be revealed; based on case 2, special boundary values of Moran's index can be found. These two sets of boundary values are neither related nor opposite to each other.

The first case is to find a spatial weight matrix to meet the extremal condition for any standardized nonzero real vector. According to equation (13) or equation (14), the condition is as below:

$$\mathbf{W} - \lambda\mathbf{E} = \mathbf{O}. \tag{15}$$

Then $\partial L(\mathbf{z})/\partial \mathbf{z} = \mathbf{O}$, $\partial L(\mathbf{z}^T)/\partial \mathbf{z}^T = \mathbf{O}^T$. Equation (15) includes a special spatial weight matrix and can be expanded as

$$\mathbf{W} = \lambda\mathbf{E} = \lambda \begin{bmatrix} 1 & 0 & \cdots & 0 \\ 0 & 1 & \cdots & 0 \\ \cdots & \cdots & \ddots & \cdots \\ 0 & 0 & \cdots & 1 \end{bmatrix} = \begin{bmatrix} \lambda & 0 & \cdots & 0 \\ 0 & \lambda & \cdots & 0 \\ \cdots & \cdots & \ddots & \cdots \\ 0 & 0 & \cdots & \lambda \end{bmatrix}, \tag{16}$$

which will be specifically explained and discussed later. If $\lambda > 0$, the maximum Moran's index is

$$I_{max} = \mathbf{z}^T(\lambda\mathbf{E})\mathbf{z} = \lambda n. \tag{17}$$

Since $\mathbf{W}$ is a globally normalized weight matrix, the sum of the elements in $\mathbf{W}$ is 1. In this instance, we have

$$\lambda = \frac{1}{n}. \tag{18}$$

Otherwise, $\mathbf{W}$ cannot meet the normalization condition. Consequently, the maximum value of Moran's index is

$$I_{max} = \frac{1}{n}n = 1. \tag{19}$$

In this case, the spatial contiguity matrix is an identity matrix. On the other hand, the basic property of correlation function is symmetry. If the maximum value is 1, the minimum value should be -1. If $\lambda<0$, we have $\lambda=-1/n$. Then the minimum Moran's index is

$$I_{\min} = -\frac{1}{n}n = -1. \tag{20}$$

In this case, the spatial contiguity matrix is a negative identity matrix.

As we know, in spatial autocorrelation theory, spatial contiguity matrix and identity matrix are incompatible with each other. One can be treated as a complement set of the other. The diagonal elements in a spatial contiguity matrix are all zero. In contrast, an identity matrix is a scalar matrix in which all of the diagonal elements are unity but other elements are all zero. Spatial autocorrelation measures indirect correlation between different geographical elements in a region. On the contrary, an identity matrix can be used to define the direct correlation between an elements and itself, that is, self-correlation. If and only if all the elements in a geographical region become directly related to themselves, and indirect correlation does not exist, the correlation strength reaches the extreme. In theory, a spatial adjacent matrix can never become an identity matrix, let alone a negative identity matrix. This implies that Moran's index values can never reach 1 or -1. It's actual value comes between -1 and 1 for ever.

The second case is to find a set of eigenvectors to meet the extremal condition for any globally normalized spatial weight matrix. According to equation (13) or equation (14), the conditions are as follows

$$\mathbf{W} - \lambda\mathbf{E} \neq \mathbf{O}. \tag{21}$$

If $\partial L(\mathbf{z})/\partial \mathbf{z}=\mathbf{O}$ and $\partial L(\mathbf{z}^T)/\partial \mathbf{z}^T=\mathbf{O}^T$ as given, then special vector $\mathbf{z}^*$ should be found to satisfy the following relation

$$(\mathbf{W} - \lambda\mathbf{E})\mathbf{z}^* = \mathbf{O}. \tag{22}$$

In this instance, the vector $\mathbf{z}^*$ is not the given vector coming from the real world. Instead, it is one of possible vector proceeding from the mathematical world and is termed eigenvector. The condition that equation (9) has a non-zero solution is $|\mathbf{W}-\lambda\mathbf{E}|=0$, which can be expressed equivalently as

$$\left|\lambda\mathbf{E} - \mathbf{W}\right| = 0. \tag{23}$$

Since $w_{ii}=w_{jj}=0$, the determinant $|\lambda\mathbf{E} -\mathbf{W}|=0$ can be expanded as

$$\lambda \begin{bmatrix} 1 & 0 & \cdots & 0 \\ 0 & 1 & \cdots & 0 \\ \cdots & \cdots & \ddots & \cdots \\ 0 & 0 & \cdots & 1 \end{bmatrix} - \begin{bmatrix} w_{11} & w_{12} & \cdots & w_{1n} \\ w_{21} & w_{22} & \cdots & w_{2n} \\ \cdots & \cdots & \ddots & \cdots \\ w_{n1} & w_{n2} & \cdots & w_{nn} \end{bmatrix} = \begin{bmatrix} \lambda & -w_{12} & \cdots & -w_{1n} \\ -w_{21} & \lambda & \cdots & -w_{2n} \\ \cdots & \cdots & \ddots & \cdots \\ -w_{n1} & -w_{n2} & \cdots & \lambda \end{bmatrix}. \quad (24)$$

The determinant can be further expanded into a polynomial of order $n$. By solving this polynomial, we can find $n$ characteristic roots $\lambda_i$, which represents $n$ eigenvalues of the spatial weight matrix ($i$=1, 2,…, $n$). Then, with the help of the approach of finding basic system of solutions, we can find $n$ eigenvectors, $\mathbf{a}_i$, corresponding to the $n$ characteristic roots, $\lambda_i$. The eigenvectors are unitized and arranged orderly to form the abovementioned orthogonal matrix, that is, $\mathbf{C}=[\mathbf{a}_1,\mathbf{a}_2,…,\mathbf{a}_n]$. These eigenvectors satisfy the following relations

$$\lambda_i = \mathbf{a}_i^\mathrm{T} \mathbf{W} \mathbf{a}_i, \quad (25)$$

$$n = \mathbf{a}_i^\mathrm{T} \mathbf{a}_i. \quad (26)$$

Equation (25) shows that the eigenvector $\mathbf{a}_i$ is the orthogonal decomposition result of $\mathbf{z}$, while equation (26) shows that the eigenvector satisfies the constraints displayed in equation (11). It is easy to verify the following equation

$$\mathbf{z} = \sum_{i=1}^{n} (\mathbf{z}^\mathrm{T} \mathbf{a}_i) \mathbf{a}_i = \sum_{i=1}^{n} (\mathbf{a}_i^\mathrm{T} \mathbf{z}) \mathbf{a}_i = \sqrt{n} \sum_{i=1}^{n} \cos(\mathbf{a}_i, \mathbf{z}) \mathbf{a}_i, \quad (27)$$

in which the cosine is defined as

$$\frac{1}{\sqrt{n}} \mathbf{z}^\mathrm{T} \mathbf{a}_i = \frac{1}{\sqrt{n}} \mathbf{a}_i^\mathrm{T} \mathbf{z} = \cos(\mathbf{a}_i, \mathbf{z}). \quad (20)$$

This implies that the inner product of $\mathbf{z}$ and $\mathbf{a}$ is the cosine of the angle between $\mathbf{z}$ and eigenvector $\mathbf{a}$, and its value reflects the correlation between $\mathbf{z}$ and eigenvector $\mathbf{a}$. Because that $\mathbf{W}$-$\lambda\mathbf{E}$ is not a zero matrix, only a specific vector $\mathbf{z}$ can be found to satisfy the above extreme condition, that is, d$L(\mathbf{z})$/d$\mathbf{z}$=0. In this case, it is impossible to guarantee that $\mathbf{z}$ has practical significance. It may only have significance in the mathematical world, but not in the real world.

### 2.3 Intersection between two sets of boundary values

In terms of the above mathematical reasoning, Moran's index has two types of boundary values. The first type of bounds are determined by both spatial weight matrix $\mathbf{W}$ and size vector $\mathbf{z}$. In theory, Moran's index value varies from -1 to 1, that is, -1≤ $I$ ≤ 1. However, in practice, Moran's index is greater than -1 but less than 1, that is, $-1 < I < 1$. The set of boundary values are absolute and correspond to the bounds of other correlation coefficients such as Pearson correlation coefficient and temporal autocorrelation coefficient. The second type of boundary values are only determined by spatial weight matrix. The upper limit is the *n* times of the maximum eigenvalue of spatial weight matrix, while the lower limit is the *n* times of the minimum eigenvalue of the spatial weight matrix. In short, one *n*th of the Moran's index varies between the minimum and maximum eigenvalues of the spatial weight matrix (Table 1).

Table 1 A comparison between two types of boundary values of Moran's index

| Boundary values | Size variable z | Contiguity matrix V | Determinant factor | Eigenvector |
|---|---|---|---|---|
| -1, 1 | Arbitrary given vector | Effective spatial contiguity matrix | Size variable and spatial weight matrix | Known, coming from reality |
| $n\lambda_{min}$, $n\lambda_{max}$ | Special vectors—eigenvector | Arbitrary spatial contiguity matrix | Spatial weight matrix | Unknown, solution to the eigen equation |

These two sets of boundary values are not incompatible with each other, but act at the same time. Obviously, the constraints of the second set of boundary values are less and thus the bounds are more relaxed in theory. The intersections of the two sets of boundary values give four sets of bounds for Moran's index (Table 2).

Table 2 Four possible intersections for the boundary values of Moran's index

| | $\lambda_{max} > 1/n$ | $\lambda_{max} < 1/n$ |
|---|---|---|
| $\lambda_{min} < -1/n$ | $-1 < I < 1$ | $-1 < I < n\lambda_{max}$ |
| $\lambda_{min} > -1/n$ | $n\lambda_{min} < I < 1$ | $n\lambda_{min} < I < n\lambda_{max}$ |

## 3 Empirical evidence

### 3.1 Study area and data

A simple case analysis is made to verify the main results derived above. The study area is Beijing, Tianjin and Hebei (BTH) region of China. The area is also termed Jing-Jin-Ji (JJJ) region in literature (Long and Chen, 2019). It contains Beijing city, Tianjin city, and the main cities of Hebei Province. There are 13 prefecture level and above cities in this region, thus the number of cities is $n = 13$. The spatial sample is very small, so it is suitable to make a teaching case. The size variable is the city population of the fifth census in 2000 and the sixth census in 2010 (Table 3). The related size measure is the nighttime light intensity in the corresponding years (Long and Chen, 2019). The spatial weight matrix is determined by the traffic mileage between cities (Table 4). The spatial weight function adopts the inverse distance function. This function is in fact the intersection of power law and hyperbolic function. Thus, the spatial contiguity is defined as

$$v_{ij} = \begin{cases} 1/d_{ij}, & i \ne j \\ 0, & i = j \end{cases}, \qquad (21)$$

in which $d_{ij}$ denotes the road distance between city $i$ and city $j$. Using equation (21), we can convert the traffic mileage matrix (**U**) into a spatial contiguity matrix (**V**), and then by global normalization, we can change spatial contiguity matrix (**V**) to the globally normalized weight matrix (**W**).

**Table 3 Population size and nighttime light intensity of cities in Beijing-Tianjin-Hebei region**

| City | 2000 | | 2010 | |
| --- | --- | --- | --- | --- |
| | City population | Nighttime light intensity | City population | Nighttime light intensity |
| Beijing | 949.6688 | 340.7774 | 1555.2378 | 439.0930 |
| Tianjin | 531.3702 | 211.2071 | 885.6234 | 297.0876 |
| Shijiazhuang | 193.0579 | 68.6254 | 275.6871 | 83.2512 |
| Tanshan | 140.3887 | 63.3617 | 163.7579 | 112.1651 |
| Qinhuangdao | 70.7267 | 23.0483 | 95.1872 | 33.1156 |
| Handan | 107.1068 | 95.1528 | 111.7417 | 108.6665 |
| Xingtai | 53.6282 | 28.1144 | 63.7797 | 35.2086 |
| Baoding | 90.2496 | 43.7416 | 98.0177 | 61.2349 |
| Zhangjiakou | 79.6580 | 27.6160 | 90.0218 | 38.6023 |
| Chengde | 32.5821 | 6.9849 | 49.8293 | 15.6265 |
| Cangzhou | 44.3561 | 41.6252 | 48.9701 | 52.6133 |
| Langfang | 29.5879 | 48.1902 | 46.6539 | 73.2010 |
| Hengshui | 24.5229 | 13.2241 | 38.2976 | 18.3238 |
| **Mean** | **180.5311** | 77.8207 | **270.9850** | 105.2453 |
| σ | **256.5845** | 91.1733 | **429.9496** | 119.1866 |

**Table 4 Road distance matrix of Beijing-Tianjin-Hebei cities based on traffic mileage**

| City | Beijing | Tianjin | Shijiazhuang | Tanshan | Qinhuangdao | Handan | Xingtai | Baoding | Zhangjiakou | Chengde | Cangzhou | Langfang | Hengshui |
| --- | --- | --- | --- | --- | --- | --- | --- | --- | --- | --- | --- | --- | --- |
| Beijing | 0 | 160.8855 | 321.7625 | 185.4770 | 288.9055 | 479.9810 | 430.2520 | 187.1300 | 198.1975 | 194.5940 | 233.4440 | 83.2755 | 299.7580 |
| Tianjin | 160.8855 | 0 | 344.5825 | 101.4105 | 242.6355 | 454.8400 | 425.3890 | 201.9420 | 332.9375 | 280.6470 | 138.6135 | 86.1555 | 259.8555 |
| Shijiazhuang | 321.7625 | 344.5825 | 0 | 423.7510 | 568.1560 | 167.2815 | 114.0840 | 138.9090 | 430.8215 | 506.6400 | 221.7565 | 283.2495 | 142.5935 |
| Tanshan | 185.4770 | 101.4105 | 423.7510 | 0 | 151.3880 | 547.4205 | 517.8910 | 289.5120 | 376.8000 | 185.3500 | 215.0285 | 144.6130 | 352.4360 |
| Qinhuangdao | 288.9055 | 242.6355 | 568.1560 | 151.3880 | 0 | 711.7120 | 662.2960 | 433.9170 | 481.3360 | 222.2030 | 375.5205 | 292.9180 | 508.4835 |
| Handan | 479.9810 | 454.8400 | 167.2815 | 547.4205 | 711.7120 | 0 | 53.4600 | 296.7465 | 606.6940 | 664.8585 | 335.0465 | 440.4685 | 214.2995 |
| Xingtai | 430.2520 | 425.3890 | 114.0840 | 517.8910 | 662.2960 | 53.4600 | 0 | 245.8830 | 557.3515 | 615.1295 | 299.4430 | 391.1260 | 167.0325 |
| Baoding | 187.1300 | 201.9420 | 138.9090 | 289.5120 | 433.9170 | 296.7465 | 245.8830 | 0 | 278.0950 | 372.0075 | 150.5130 | 147.8300 | 144.8405 |
| Zhangjiakou | 198.1975 | 332.9375 | 430.8215 | 376.8000 | 481.3360 | 606.6940 | 557.3515 | 278.0950 | 0 | 372.8730 | 411.7425 | 257.5700 | 455.2955 |
| Chengde | 194.5940 | 280.6470 | 506.6400 | 185.3500 | 222.2030 | 664.8585 | 615.1295 | 372.0075 | 372.8730 | 0 | 407.1040 | 259.8085 | 495.3555 |
| Cangzhou | 233.4440 | 138.6135 | 221.7565 | 215.0285 | 375.5205 | 335.0465 | 299.4430 | 150.5130 | 411.7425 | 407.1040 | 0 | 149.7245 | 140.0620 |
| Langfang | 83.2755 | 86.1555 | 283.2495 | 144.6130 | 292.9180 | 440.4685 | 391.1260 | 147.8300 | 257.5700 | 259.8085 | 149.7245 | 0 | 237.8790 |
| Hengshui | 299.7580 | 259.8555 | 142.5935 | 352.4360 | 508.4835 | 214.2995 | 167.0325 | 144.8405 | 455.2955 | 495.3555 | 140.0620 | 237.8790 | 0 |

## 3.2 Calculation results

Absolute boundary values cannot be verified by examples, while relative boundary values can be illustrated by case studies. First, we can calculate the eigenvalues and the corresponding eigenvectors of the normalized spatial weight matrix. The inner product of each eigenvector $\mathbf{a}_i$ and the size vector $\mathbf{z}$ can be worked out conveniently (Table 5). Multiplying the eigenvalues of $\mathbf{W}$ by $n$ yields the eigenvalues of $n\mathbf{W}$. If the $i$th eigenvalue of $\mathbf{W}$ is $\lambda_i$, then the corresponding eigenvalue of $n\mathbf{W}$ is $n\lambda_i$. The maximum and minimum eigenvalues of $\mathbf{W}$ are $\lambda_{max}=0.0816$ and $\lambda_{min}=-0.0304$. So, the maximum and minimum eigenvalues of $n\mathbf{W}$ are $n\lambda_{max}=1.0604$ and $n\lambda_{min}=-0.3947$. Moran's index comes between the maximum eigenvalue and the minimum eigenvalue of $n\mathbf{W}$. Using the data displayed in Table 3 and the results shown in Table 5, it is easy to testify equation (27).

**Table 5 Eigenvectors and corresponding eigenvalues of normalized spatial weight matrix of Beijing-Tianjin-Hebei cities based on road distances**

| City | Vector1 | Vector2 | Vector3 | Vector4 | Vector5 | Vector6 | Vector7 | Vector8 | Vector9 | Vector10 | Vector11 | Vector12 | Vector13 |
|---|---|---|---|---|---|---|---|---|---|---|---|---|---|
| Beijing | -0.5842 | -0.0131 | -0.3908 | -0.0072 | -0.0619 | -0.3718 | 0.0557 | -0.0246 | -0.0553 | -0.2602 | 0.2791 | 0.2599 | -0.3791 |
| Tianjin | -0.3332 | -0.0094 | 0.6780 | -0.0441 | -0.0729 | 0.3774 | -0.0664 | -0.0766 | 0.2758 | -0.0032 | 0.0851 | 0.2372 | -0.3602 |
| Shijiazhuang | -0.0019 | 0.1559 | 0.0143 | -0.7197 | -0.1673 | -0.0310 | -0.0309 | -0.2891 | -0.3144 | 0.2816 | 0.0509 | -0.3127 | -0.2559 |
| Tangshan | 0.0875 | 0.0005 | -0.4403 | 0.0437 | -0.4886 | 0.3809 | -0.0813 | -0.1190 | 0.1723 | -0.0212 | -0.5174 | 0.1589 | -0.2622 |
| Qinhuangdao | 0.0133 | -0.0003 | 0.1925 | -0.0473 | 0.3450 | -0.5252 | -0.2744 | -0.1775 | 0.1016 | -0.0293 | -0.6358 | 0.0955 | -0.1853 |
| Handan | -0.0172 | 0.6748 | 0.0006 | 0.1962 | 0.0385 | 0.0105 | 0.0149 | 0.1134 | 0.2272 | -0.2941 | -0.0053 | -0.5466 | -0.2329 |
| Xingtai | 0.0189 | -0.7160 | -0.0155 | -0.0220 | 0.0495 | 0.0428 | -0.0001 | 0.0227 | 0.1503 | -0.2845 | 0.0073 | -0.5629 | -0.2468 |
| Baoding | 0.0708 | -0.0142 | -0.0042 | 0.5627 | 0.1753 | 0.1456 | -0.0467 | -0.5614 | -0.4497 | 0.1436 | 0.0985 | -0.0719 | -0.2666 |
| Zhuangjiakou | 0.0319 | 0.0008 | 0.0593 | 0.0107 | -0.0056 | 0.1442 | -0.5478 | 0.5627 | -0.5436 | -0.1590 | -0.0448 | 0.0454 | -0.1814 |

| | | | | | | | | | | | | |
|---|---|---|---|---|---|---|---|---|---|---|---|---|
| **Chengde** | 0.0179 | 0.0022 | 0.1431 | -0.0408 | 0.1026 | 0.0712 | 0.7711 | 0.2202 | -0.3706 | -0.1657 | -0.3360 | 0.0806 | -0.1861 |
| **Cangzhou** | -0.0314 | 0.0042 | -0.3323 | -0.0845 | 0.6147 | 0.2683 | 0.0284 | 0.2578 | 0.2211 | 0.4774 | 0.0366 | 0.0370 | -0.2913 |
| **Langfang** | 0.7291 | 0.0232 | 0.0234 | -0.0912 | -0.0155 | -0.1653 | 0.0381 | -0.0068 | 0.1448 | -0.2000 | 0.3381 | 0.3043 | -0.4026 |
| **Hengshui** | 0.0148 | -0.0809 | 0.1460 | 0.3208 | -0.4263 | -0.3862 | 0.1008 | 0.3226 | 0.0535 | 0.5857 | 0.0045 | -0.1437 | -0.2395 |
| **Eigenvalue** | -0.0304 | -0.0276 | -0.0160 | -0.0135 | -0.0122 | -0.0096 | -0.0050 | -0.0031 | -0.0001 | 0.0020 | 0.0053 | 0.0286 | 0.0816 |
| **$a_i^T z$** | -2.6997 | 0.1533 | -0.2803 | -0.4240 | -0.5768 | -0.4183 | -0.1109 | -0.6212 | 0.3368 | -0.9143 | 1.2619 | 1.3071 | -0.5228 |

Second, let's compute Moran's index based on different measures, including urban population size and nighttime light number. All the values of Moran's index are less than 0, and the $P$ values of them are all great than 0.1. This indicates that the spatial autocorrelation of the cities in Beijing-Tianjin-Hebei region are not significant. The aim of this research is not at the spatial analysis of the cities in the study area. What we focus on is the boundary values of Moran's index. It can be seen that all Moran's index values are between -1 and 1, and all Moran's index values are between the maximum and minimum eigenvalue of $n\mathbf{W}$. That is, $-0.3947 < I < 1.0604$ (Table 6).

**Table 6 Moran's index values based on different measures of Beijing-Tianjin-Hebei cities**

| Measure | Year | Moran's $I$ | P-value (with constant) | P-value (no constant) |
|---|---|---|---|---|
| **Population** | 2000 | -0.1513 | 0.1251 | 0.1379 |
| | 2010 | -0.1394 | 0.1625 | 0.1769 |
| **Nighttime light number** | 2000 | -0.1221 | 0.1962 | 0.1379 |
| | 2010 | -0.0962 | 0.3206 | 0.3509 |

## 4 Discussion

As a spatial statistical measurement, Moran's index is actually a spatial autocorrelation coefficient defined in a 2-dimensional space and can be expressed a real quadratic form. Two sets of boundary values of Moran's index can be derived by finding conditional extremum based on the quadratic form. First, assuming that the size vector is any real vector, we find a spatial weight matrix to maximize the absolute value of Moran index. In this case, the spatial contiguity matrix is an identity matrix, and the extremum value of the absolute value of Moran's index is 1. This suggests that Moran's index comes between -1 and 1 in theory. Second, assuming that the space weight matrix is given, we find the eigenvectors of the normalized weight matrix to satisfy the extremum condition. In this case, for $n$ geographical elements, we can find $n$ eigenvectors, corresponding to $n$ eigenvalues.

In this case, Moran's index varies from *n* times of the minimum eigenvalue and *n* times of the maximum eigenvalue. The differences between the two sets of boundary values for Moran's index lie in two aspects. On the one hand, the first set of boundary values, ±1, is a prior and is applicable to any standardized real vector. Therefore, this set of boundary values is general and practical in spatial analysis. In contrast, the second set of boundary, $n\lambda_{min}$ and $n\lambda_{max}$, is unknown beforehand and is suitable for the Moran's index based on special spatial weight matrix. A set of good boundary values is predictable in advance without calculation. On the other hand, the first set of boundary values represents the limit of autocorrelation intensity, and it is based on the correlation of a variable and itself without lag or displacement. As indicated above, it is suitable for any real vectors. In contrast, the second set of boundary values reflect the two extreme extension directions of the given size vector. The eigenvalues reflect the extension lengths in two special directions.

A measurement such as Moran's index is often an indicator that condenses a large number of data into one data point to describe the main characteristics of something. A good measurement usually has a definite threshold or boundary values. These values have two basic properties: First, they are known in advance and do not need to be calculated each time. Second, they are universal and will not be different due to different research objects. For example, Pearson correlation coefficient is a basic measurement in statistics. Its basic threshold value is 0, and its bounds are -1 and 1. Therefore, Pearson correlation coefficient is very easy to use in research work. Many measurements have more than one set of boundary value. In different cases, we need different boundary values for a measurement. Let's examine two simple examples. One is the level of urbanization, the other is the fractal dimension of urban form based on box-counting method (Table 7). The level of urbanization is defined as the ratio of urban population to total population in a geographical region. Therefore, the urbanization level has a lower limit of 0 (or 0%) and an upper limit of 1 (or 100%). For many countries, the curve of urbanization level can be modeled by logistic function (Cadwallader, 1996; Chen, 2009; Pacione, 2009; Rao *et al*, 1989; United Nations, 2004; Zhou, 1995). According to the logistic model of urbanization curve, we can obtain another set of boundary values, that is initial value $L_0>0$ and capacity value $L_{max}<1$. Fractal dimension of urban form can be treated as a scaling exponent of urban land use patterns (Batty and Longley, 1994; Benguigui *et al*, 2000; Frankhauser, 1998; Shen, 2002). According to the definition of fractals, a fractal dimension value is greater than the fractal's topological dimension ($d_T$), while according to the property of box-counting method, a fractal dimension value is less than the Euclidean dimension ($d_E$) of the embedding space of the fractal. This suggests the first set of bounds of fractal dimension (Chen, 2011; Mandelbrot, 1982). Using box-counting method, we can obtain a multifractal dimension spectrum, and the generalized correlation dimension values come between the fractal dimension for the moment order of positive infinity ($D_{+\infty}$) and the fractal dimension for the moment order of minus infinity ($D_{-\infty}$). This suggests the second set of bounds of fractal dimension (Chen, 2014; Feder, 1988). A time series of fractal dimension of urban growth can be modeled by Boltzmann equation, which gives the minimum fractal dimension ($D_{min}$) and the maximum fractal dimension ($D_{max}$). This suggests the third set of bounds of fractal dimension (Chen, 2011; Chen, 2018). Urbanization level and fractal dimension have no direct relationship with Moran's index. However, through the analogy between Moran's index and urbanization level as well as fractal dimension, we can better understand why Moran's index has at least two sets of boundary values at the same time.

**Table 7 Boundary values of urbanization level and box-counting fractal dimension of urban form**

| Measure | Criterion | Boundary values | |
|---|---|---|---|
| | | Lower limit | Upper limit |
| Level of urbanization | Definition | 0 | 1 (100%) |
| | Logistic model | $L_0 > 0$ | $L_{max} < 1$ |
| Fractal dimension of urban form | Spatial measurement | Topological dimension $d_T=0$ | Euclidean dimension of embedding space $d_E=2$ |
| | Boltzmann equation | Minimum value $D_{min} > 0$ | Maximum value $D_{max} < 2$ |
| | Multifractal spectrum | $D_{+\infty} > 0$ | $D_{-\infty}$ |

In order to clarify the general boundary values of Moran's index, let's draw a comparison between Moran's index and the autocorrelation function of time series. As indicated above, Moran's index in a spatial autocorrelation coefficients, and a temporal autocorrelation function consists of a series of temporal correlation coefficients. Given a time lag, we have a temporal autocorrelation coefficient. In a time autocorrelation coefficient set, no absolute value of the autocorrelation coefficient value can exceed 1. Let $\rho$ represent temporal autocorrelation function. The main similarities and differences can be summarized as follows. First, both temporal autocorrelation coefficient and Moran's index can be expressed by as quadratic forms. Based on standardized vector and globally normalized weight matrix, the quadratic form of time autocorrelation function is similar to the quadratic form of spatial autocorrelation coefficient. Second, the extremum value of temporal autocorrelation function and Moran's index is 1. In the case, both temporal contiguity matrix and spatial contiguity matrix are identity matrix. For time autocorrelation, if and only if time lag equals 0, a temporal autocorrelation coefficient equals 1, or else the autocorrelation coefficient comes between -1 and 1. If $\rho>0$, the maximum value is 1, but if $\rho<0$, the minimum value is greater than -1. For spatial autocorrelation, if and only if space displacement is 0, Moran's index equals 1, or else Moran's index comes between -1 and 1. If $I>0$, the maximum value is 1, but if $I<0$, the minimum value is greater than -1. No matter for time autocorrelation or spatial autocorrelation, only the lag or displacement is 0, and the autocorrelation coefficient reaches the maximum, i.e. 1, and in this instance, the time or spatial contiguity matrix, **V**, is a unit matrix, **E** (**V**=**E**). Since the autocorrelation coefficient given by the identity matrix is 1 instead of -1, the minimum value of the autocorrelation coefficient cannot reach -1. On the other hand, the autocorrelation coefficient based on zero lag or zero displacement is known, and there is no information, so it is usually not considered in practical applications. Third, the temporal autocorrelation is described by a function, while spatial autocorrelation is characterized by a parameter. The temporal autocorrelation function is a set comprising a series of numerical values, while the spatial autocorrelation coefficient is only one numerical value. The latter can be regarded as a set of single element. For simplicity, let time lag $\tau=1$, we have the basic temporal autocorrelation coefficient, which is most comparable with Moran's index. The above simple comparison between temporal and spatial autocorrelation coefficients helps us understand the boundary values of Moran's index readily (Table 8).

**Table 8 Quadratic forms and bounds of temporal autocorrelation function and Moran's index**

| Item | Temporal autocorrelation function | Spatial autocorrelation coefficient: Moran's index |
|---|---|---|
| Formula | $\rho_\tau = \mathbf{x}^T(\frac{1}{n}\mathbf{V}_\tau)\mathbf{x}$ | $I = \mathbf{z}^T\mathbf{W}\mathbf{z} = \mathbf{z}^T(\frac{1}{V_0}\mathbf{V})\mathbf{z}$ |
| Extremum condition | No time lag | No spatial displacement |
| Spatial contiguity matrix for extreme value | Identity matrix (**V**=**E**) | Identity matrix (**V**=**E**) |
| Whether the extreme values can be reached | No | No |
| Theoretical value range | $-1 \leq \rho \leq 1$ | $-1 \leq I \leq 1$ |

| Actual value range | $-1 < \rho < 1$ | $-1 < I < 1$ |

The novelty of this work lies in two aspects. Firstly, two sets of boundary values of Moran's index are derived from new points of view. Secondly, the association and differences between the two sets of boundary values are clarified. The main shortcomings of this research are as follows. First, the demonstration process of the lower limit of Moran's index is based on symmetry idea of correlation function rather than direct mathematical derivation. By means of any standardized real vector and globally normalized matrix, it is easy to derive the upper limit of Moran's index by conditional extreme values. However, it is difficult to derive the lower limit directly by using the similar method. Second, in the process of mathematical derivation for the boundary values, the properties of globally normalized symmetric matrix are not taken into account totally. In the derivation process based on eigenvalues, the symmetry of spatial weight matrix was considered. However, the standardized size vector and normalization of weight matrix are not taken into consideration. In the derivation process based on conditional extreme values, the basic property of distance matrix are not considered. If a spatial contiguity matrix violates the distance axiom, the calculated results of the boundary values of Moran's index may exceed plus or minus 1.

## 5 Conclusions

A good measurement index in scientific research has clear boundary values or threshold value. The boundary values of an index are often more than one set. The general boundary values for a measurement index are known numbers, which can be determined a prior. By using the conditional extreme value method based on quadratic forms, we can derive two sets of bounds values for Moran's index. Different sets of boundary values bears different uses in spatial analysis. *First, the general or absolute boundary values of Moran's index are* -1 *and* 1. These boundary values are a prior and determined by both size vector and spatial weight matrix. They are suitable for any spatial autocorrelation coefficient values in geographical spatial analysis. In theory, we have -1≤ Moran's $I$ ≤1, but in practice, we have we have -1< Moran's $I$ <1. In short, for the absolute value of Moran index, 1 is an unattainable value. Only when $n$ geographical elements are self-correlated without spatial displacement and time lag can the absolute value of Moran index reach 1. However, the definition of spatial weight matrix limits the Moran index, making it unable to reach the extreme values. In practical work, if the absolute value of the Moran index exceeds 1, it may be due to the problem of constructing the spatial weight matrix. *Second, the special or relative boundary values of Moran's index are determined by the minimum and maximum eigenvalues of spatial weight matrix ($\lambda_{min}$, $\lambda_{max}$).* Based on globally normalized spatial weight matrix, Moran's index comes between $n\lambda_{min}$ and $n\lambda_{max}$. These boundary values are determined by spatial weight matrix and suitable for the spatial autocorrelation coefficient values in special geographical region. Compared with the first set of boundary values, the second boundary values are unknown numbers. Therefore, the theoretical significance of the second set of boundary values is higher than the practical significance. Only for specific research needs, the second set of boundary values has application value and practical significance.

## Acknowledgement:

This research was sponsored by the National Natural Science Foundation of China (Grant No. 42171192). The support is gratefully acknowledged.

# References

Addison PS (1997). *Fractals and Chaos: An Illustrated Course*. Bristol and Philadelphia: Institute of Physics Publishing

Batty M, Longley PA (1994). *Fractal Cities: A Geometry of Form and Function*. London: Academic Press

Benguigui L, Czamanski D, Marinov M, Portugali J (2000). When and where is a city fractal? *Environment and Planning B: Planning and Design*, 27(4): 507–519

Cadwallader MT (1996). *Urban Geography: An Analytical Approach*. Upper Saddle River, NJ: Prentice Hall

Chen YG (2009). Spatial interaction creates period-doubling bifurcation and chaos of urbanization. *Chaos, Solitons & Fractals*, 42(3): 1316-1325

Chen YG (2011). Fractal systems of central places based on intermittency of space-filling. *Chaos, Solitons & Fractals*, 44(8): 619-632

Chen YG (2014). Multifractals of central place systems: models, dimension spectrums, and empirical analysis. *Physica A: Statistical Mechanics and its Applications*, 402: 266-282

Chen YG (2018). Logistic models of fractal dimension growth of urban morphology. *Fractals*, 26(3): 1850033

de Jong P, Sprenger C, van Veen F (1984). on extreme values of Moran's I and Geary's C. *Geographical Analysis*, 16(1): 985-999

Feder J (1988). *Fractals*. New York: Plenum Press, 1988

Frankhauser P (1998). The fractal approach: A new tool for the spatial Analysis of urban agglomerations. *Population: An English Selection*, 10(1): 205-240

Haggett P, Cliff AD, Frey A (1977). *Locational Analysis in Human Geography*. London: Edward Arnold

Long YQ, Chen YG (2019). Multi-scaling allometric analysis of the Beijing-Tianjin-Hebei urban system based on nighttime light data. *Progress in Geography*, 38(1): 88-100 [In Chinese]

Mandelbrot BB (1983). *The Fractal Geometry of Nature.* New York: W. H. Freeman and Company

Moran PAP (1950). Notes on continuous stochastic phenomena. *Biometrika*, 37(1-2): 17-33

Odland J (1988). *Spatial Autocorrelation*. London: SAGE Publications

Pacione M (2009). *Urban Geography: a Global Perspective (Third edition)*. London: Taylor & Francis

Rao DN, Karmeshu Jain VP (1989). Dynamics of urbanization: the empirical validation of the replacement hypothesis. *Environment and Planning B: Planning and Design*, 16(3): 289-295

Shen GQ (2002). Fractal dimension and fractal growth of urbanized areas. International Journal of Geographical Information Science, 16(5): 419-437

Tiefelsdorf M, Boots B (1995). The exact distribution of Moran's *I*. *Environment and Planning A*, 27(6): 985-999

United Nations (2004). *World Urbanization Prospects: The 2003 Revision*. New York: U.N. Department of Economic and Social Affairs, Population Division

Xu F (2021). Improving spatial autocorrelation statistics based on Moran's index and spectral graph theory. *Urban Development Studies*, 28(12): 94-103 [In Chinese]

Zhou YX (1995). *Urban Geography*. Beijing: Commercial Press